# Reliable Link-Based Routing Protocol for Highly Dynamic Mobile Adhoc Networks

Sahaya Rose Vigita.E[#1], Golden Julie.E[*2]

[#]*Final Year ME, Dept. of CSE, Regional Centre of Anna University, Tirunelveli, India*
[*]*Assistant Professor, Dept. of CSE, Regional Centre of Anna University, Tirunelveli, India*

*Abstract*— **Traditional topology-based MANET routing protocols use stateful routing which increases the processing, communication and memory overheads. The high mobility of nodes in MANETs makes it difficult to maintain a deterministic route. To overcome this, stateless geographic routing protocols which ensure reliable data delivery have been proposed. It is found that link instability can be a major factor for unreliable data delivery. Driven by this issue, Link and Position based Opportunistic Routing (L-POR) protocol which chooses a forwarder based on the reception power of a node has been proposed. A back-up scheme is also proposed to handle communication holes. Simulation results show that the proposed protocol achieves excellent performance even under high node mobility.**

*Keywords*- **Mobile Adhoc Networks, reliability, opportunistic routing, mobility, link stability.**

## I. INTRODUCTION

Traditional routing protocols in mobile adhoc networks (MANETs) introduce probe packets to store the cost of the path that they traverse and enable route discovery. Examples of costs are the number of hops in the path, the probability of packet loss, the estimated delay along the path, etc. Based on the cost information an appropriate path is chosen. The selected paths are stored in a routing table. These protocols may fail due to network disconnectivity caused by node mobility, node sparseness or propagation variations. To overcome these problems Opportunistic routing protocols have been proposed.

Opportunistic routing [1] utilises the broadcast nature of the wireless medium to choose at least one relay node as a forwarder at each hop in a highly dynamic environment. The neighbouring nodes are prioritized according to some metrics such as distance to the destination, link stability etc. The node with the highest priority is chosen as the best forwarder and is specified in the next hop field of the packet. When the packets are broadcast the nodes other than the best forwarder receive the packet by eavesdropping. Among the receivers, the node specified as the best forwarder in the packet header becomes the next forwarder. Other nodes get suppressed when they hear the transmission by the best forwarder.

Routing information becomes stale due to the unpredicted nature of the wireless environment. Hence Geographic routing [2] is used to exploit the one-hop neighbour's geographic information. This information helps to gradually approach and eventually reach the destination in a hop-by-hop fashion. No energy is spent on route discovery [3]. Memory overheads due to routing table maintenance are reduced. Traffic caused by route queries and replies are reduced.

The forwarding strategy can fail when no forwarders with a positive progress towards destination are found. In such a situation, a routing hole is said to be encountered [2]. Holes may be induced by obstacles, unreliable nodes, the boundaries of a wireless network and the like. Hence, a back-up mode algorithm is required to enable routing around the hole in an effective and efficient manner. Traditional routing protocols may fail in mobile adhoc networks (MANETs) due to unpredictable network disconnectivity caused by node mobility. Hence Opportunistic routing protocols have been proposed. Opportunistic routing [1] selects a best forwarder at each hop according to some metrics such as distance to the destination, link stability etc. To prevent routing information from becoming stale Geographic routing [2] is used to exploit the one-hop neighbour's geographic information. A routing hole is said to be encountered when no forwarders are found. Hence, a back-up mode algorithm is required to enable routing around the hole in an effective and efficient manner.

## II. REVIEW OF LITERATURE

Several Geographic Opportunistic routing protocols that enhance reliability have been proposed earlier. Some of them are discussed below.

### A. Simple Opportunistic Adaptive Routing Protocol (SOAR)

Rozner, Seshadri et al. [2009] proposed Simple Opportunistic Adaptive Routing Protocol (SOAR) [4] which is a proactive link state routing protocol. Every node periodically measures and disseminates link quality in terms of Expected Transmission Count (ETX). Based on this information, a sender selects the default path and a list of forwarding nodes. This protocol achieves high throughput and deals efficiently with fairness. However, periodic measurement and dissemination of link quality drains node energy. Each node maintains a routing table adding to memory overhead.

### B. Utility based Opportunistic Routing

Jie Wu, Mingming Lu and Feng Li [2008] applied opportunistic routing to a utility-based routing [5] where the successful delivery of a data packet generates a benefit. The optimal route depends on the benefit value. Accordingly, an optimal centralized algorithm and an approximation distributed algorithm are proposed to solve the routing problem. Failure of one path leads to retransmissions using alternate paths. Retransmission has a negative impact on routing and it has not been evaluated.

*C. Geographic Opportunistic Routing (GOR)*

Kai Zeng et al. [2007] analysed one-hop throughput of Geographic Opportunistic Routing (GOR) using the one-hop throughput metric [6]. A local metric named Expected One-hop Throughput (EOT) to balance the benefit and cost is proposed. Based on the EOT, a local candidate selection and prioritization algorithm is proposed. The proposal fails to discuss the effect of retransmission of packets. Forwarding area to be covered is not detailed.

*D. Virtual Routing Protocol (VRP)*

Luiz Carlos P. Albini et al. [2006] proposed the Virtual Routing Protocol (VRP) [7] which is a hybrid source routing protocol. VRP defines a logical structure over the network which is unrelated to the physical network topology. Routes between units are built by translating virtual paths into physical routes. Although the protocol is found to achieve high packet delivery ratio, VRP performs poorly under heavy traffic conditions since units are not able to maintain up-to-date route information about their logical neighbours.

*E. Greedy Perimeter Stateless Routing (GPSR)*

Brad Karp and H.T. Kung [2000] proposed Greedy Perimeter Stateless Routing (GPSR) [8] protocol which makes greedy forwarding decisions using only information about a router's immediate neighbours in the network topology. When greedy forwarding is impossible, the algorithm recovers by routing around the perimeter of the region. This protocol has a few disadvantages too. Because GPSR's beacons are sent continuously, each beaconing interval results in a constant level of routing protocol traffic. The addition of location registration and lookup traffic for a location database is found to increase the overhead of GPSR.

*F. Extremely Opportunistic Routing Protocol (ExOR)*

Sanjit Biswas and Robert Morris [2005] proposed Extremely Opportunistic Routing Protocol (ExOR) [9], an integrated routing and MAC protocol that increases the throughput of large unicast transfers in multi-hop wireless networks. ExOR chooses each hop of a packet's route after the transmission for that hop. It chooses the forwarder with the lowest remaining cost to the ultimate destination. Though it transmits each packet fewer times than traditional routing causing less interference for other users of the network and of the same spectrum, the ExOR header grows with the batch size and many transfers may only have a few packets.

III. PROPOSED ROUTING PROTOCOL

Link and Position based Opportunistic Routing (L-POR) protocol is designed to achieve maximum reliability in a mobile adhoc network. It combines geographic and opportunistic routing to achieve high packet delivery ratio.

The protocol chooses the best forwarder based on the receptive power. When the best forwarder fails, a candidate node takes over the forwarding function. Trigger nodes trigger a hole handling mechanism when routing holes are encountered.

*A. Architecture design*

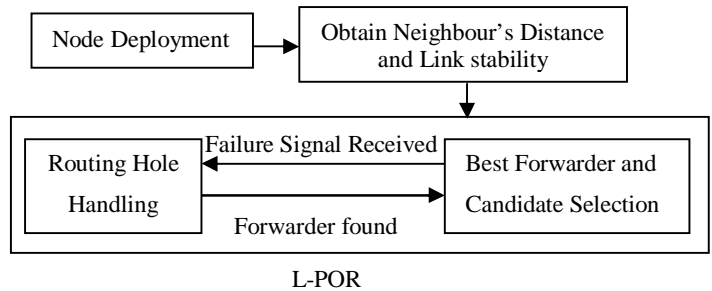

L-POR

Fig. 1 Architecture Diagram of L-POR

Fig. 1 shows the architecture design of L-POR. For randomly deployed nodes, the distance and link metrics calculations enable best forwarder and candidate selection. On receiving a forwarding failure signal, a routing hole handling module is invoked. When a potential forwarder is found the normal L-POR routing algorithm is invoked.

*B. Obtaining Distance and Link metrics*

Distance calculation between any two nodes, say node a and node b is based on Euclidean distance given by equation 1.

$$d = \sqrt{(x1 - x2)^2 + (y1 - y2)^2} \qquad (1)$$

where x1 and x2 are the x-coordinates of nodes a and b respectively and y1 and y2 are the y-coordinates of nodes a and b respectively.

Free-space propagation model can be used to predict the received strength when the transmitter and receiver have clear unobstructed line-of-sight path between them. When system losses are neglected, the free space power received by a receiver antenna separated from a transmitting antenna by a distance d is given by Friis's free-space equation [10] as,

$$P_r(d) = P_t \, G_t \, G_r \left[\frac{\lambda}{4\pi d}\right]^2 \qquad (2)$$

where,

$P_t$ = Power transmitted by the transmitter
$G_t$ = Antenna gain of transmitter
$G_r$ = Antenna gain of receiver
$\lambda$ = Wavelength
$d$ = Euclidean distance between sender and receiver

Even if no matter exists between sender and receiver, the signal still experiences free space loss due to the distance traversed. As soon as there is matter between the sender and receiver, the situation becomes more complex. Hence a system loss L [11] is included in equation 2 and modified as,

$$P_r(d) = \frac{P_t \, G_t \, G_r \, \lambda^2}{(4\pi d)^2 L} \qquad (3)$$

A link with maximum $P_r$ is considered more stable compared to its neighbors and hence reliable for packet transmission.

*C. Routing Mechanism*

The source node obtains the address of the destination from a location registration and lookup service. It then attaches destination's address to the packet header. If the destination is within the source's transmission range, then the next hop is the destination. The packets are forwarded directly and the routing process ends. Otherwise, neighbours are prioritized based on their link stability. The node which makes positive progress towards the destination and with the maximum power for reception gets the highest priority to become the best forwarder.

Forwarding area is selected as the intersection area of the transmission range of the source and half of the transmission range of the best forwarder. Among the nodes within this intersection area, only those nodes which are closer to the destination than the source and which are farther from the destination than the best forwarder, become the candidate nodes. A forwarding table consisting of the source id, destination id, best forwarder id and the ids of candidate nodes is maintained by the source for a particular period of time. The candidate list is attached to the packet header and the packet is broadcast. The best forwarder and the candidate nodes cache the packets.

The candidate nodes listen to the medium for a threshold amount of time. If the best forwarder fails to transmit the packets within this threshold time, then the candidate node with the next highest priority transmits the packet. All other candidate nodes get suppressed on hearing the transmission and drop the cached packets. Duplicate packets can be identified using a sequence number and are not propagated further.

*D. Forwarding Node Selection*

In Fig 2 node S is the source and D is the destination node. R is the radius of the transmission range of node S. The transmission range of S is denoted by the dotted circle.

The nodes in the area enclosed within the dashed arc make positive progress towards the destination. From these nodes, the one with maximum power for reception is chosen as the best forwarder, namely node B.

R/2 denotes the radius of half the transmission range of node B. The intersection area of the transmission range of S and half of the transmission range of B is taken as the forwarding area. Nodes within the forwarding area, other than node B, become candidate nodes, namely nodes H, A and F.

**Algorithm 1**: Best Forwarder and Candidate Nodes Selection

1. *Find if Destination node is in the Neighbour List.*
2. *If found, set the next hop as Destination node and exit. Else continue.*
3. *For each node in the Neighbour List, do the following:*

   //Checking for positive progress towards destination

3.a. *Check if its distance from the Destination node is greater than or equal to the distance between the current node and Destination node. If yes, break. Else, add node to an array.*
4. *Calculate the reception power for all the nodes in the array using equation 3.*
5. *Choose the node having the maximum reception power as the Best Forwarder.*

// Selecting candidate nodes

6. *For each node in the array, other than the Best forwarder, do the following:*
6.a. *Check if it is within half the radius of the Best Forwarder's transmission range.*
6.b. *If yes, check if its distance to the destination is lesser than the distance of the source node to the destination and the node's distance to destination is greater than distance of the Best Forwarder to the destination.*
6.c. *If yes, add the node to Candidate list.*
7. *Exit*

*E. Packet Header Format*

L-POR requires updating of the packet header during each hop. The packet header format consists of the fields shown in Table 1.

TABLE 1
PACKET HEADER FORMAT

| Field | Denotes |
|---|---|
| seq | Sequence Number |
| S | Source ID |
| D | Destination ID |
| F | Best Forwarder ID |
| CN1 | Candidate Node 1 ID |
| CN2 | Candidate Node 2 ID |
| st | Send Time |

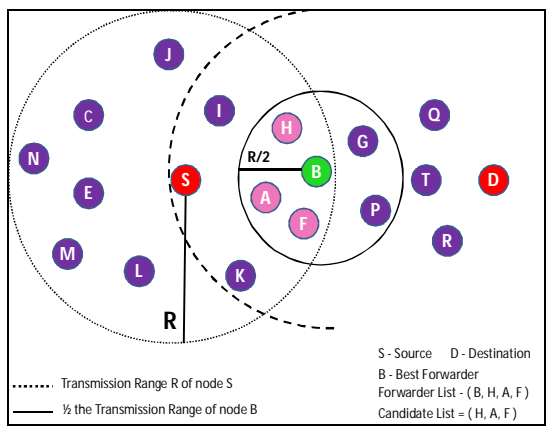

Fig. 2 Best forwarder and Candidates Selection

Sequence number is used to avoid duplicate forwarding. The destination address is attached to the header by the source node. According to [12], two candidate nodes are sufficient to provide high robustness. Hence two fields CN1 and CN2 are included in the packet header. The time of packet transmission st is also included in the header. This is used by the candidate nodes to calculate the threshold time within which the best forwarder is expected to broadcast the packets.

*F. Function of the Candidate Node*

The candidate nodes add a threshold time to the field 'st' and wait for that period. If no transmission is overheard during this period, the candidate node understands that the best forwarder has failed. The forwarding operation is then taken over by the candidate node. The candidate node now becomes the best forwarder and applies Algorithm-1 to forward the packets.

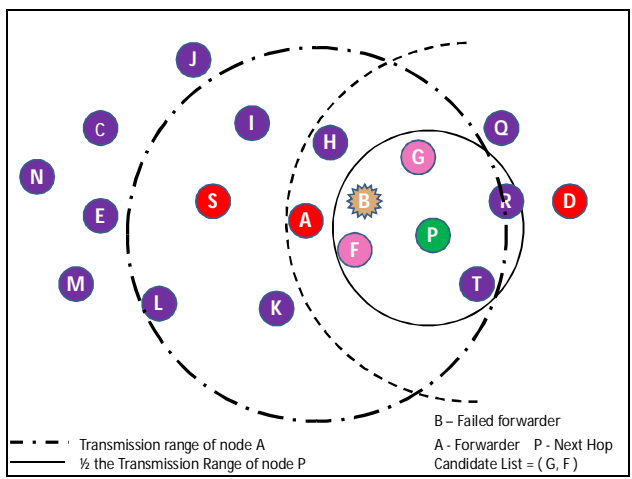

Fig. 3 Forwarding by Candidate node

In Fig. 3 the best forwarder node B fails to transmit packets. So the candidate node having the next highest priority, node A takes over the forwarding function. Node A chooses node P as the best forwarder and nodes G and F as the candidate nodes. Even though node T is within the forwarding area, it is not chosen as a candidate, because its distance to destination is lesser that the distance of node P to the destination.

*G. Structure of Forwarding Table*

Every node that is a sender or a relay node maintains a forwarding table for the packets of each flow as shown in Table 2. The table entry has an expiry time within which a transmission is expected to be completed. Thus the overhead in constructing and maintaining the forwarding table is much lesser compared to that of a traditional routing table.

TABLE 2
STRUCTURE OF FORWARDING TABLE

| (Source, Destination) | Next Hop | Candidates |
|---|---|---|
| S, D | B | A, H |
| M, Q | S | E, L |

*H. Routing Hole Handling*

Communication holes may exist since nodes are not uniformly distributed. When the best forwarder seeks the next hop node and finds none, a communication void is said to be encountered. The protocol then switches to a routing hole handling mechanism. When the best forwarder encounters a communication hole, it sends a void signal to the previous forwarder. The previous forwarder becomes the trigger node and the best forwarder becomes the void node. The trigger node triggers Algorithm-2 excluding the void node in order to avoid looping. If the next hop is the destination, packets are forwarded and an acknowledgement is sent to the trigger node. If a neighbour that makes positive progress to the real destination and which is nearer to the destination than the current node is found, then the routing switches back to the normal L-POR routing algorithm. If no forwarders are found, then the routing fails and a disrupt message is sent to the trigger node.

In Fig. 4 node A has chosen node B as the next forwarder. Node B finds no forwarders to forward the packets. In such a situation node B is said to encounter a routing hole. It sends a void warning signal to node A. Now, node B becomes the void node and node A becomes the trigger node. Node A switches to a hole handling algorithm. Node A chooses another forwarder based on Algorithm-2 excluding the void node. It may route around the void through C-H-G-T or L-O-P-R. If destination is reached, then an acknowledgement is sent to the trigger node, else a disrupt signal is sent.

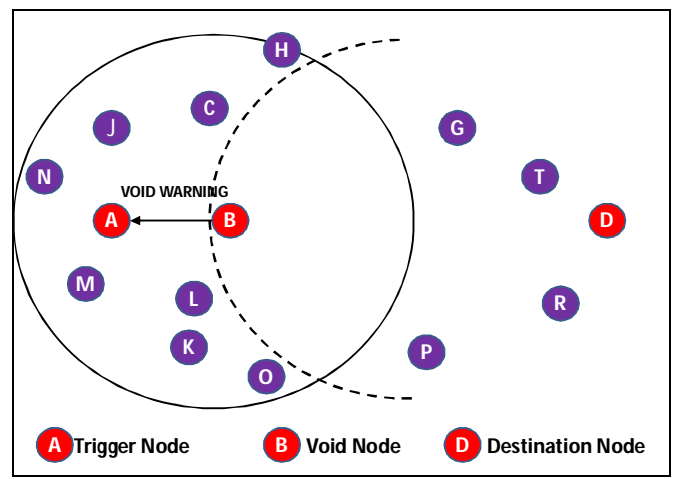

Fig. 4 Sending a void warning signal

**Algorithm 2 :** Routing Hole Handling

1. *If the current node receives void warning from the Best Forwarder, then do the following:*
1.a. *Set the Best Forwarder as Void node.*
1.b. *Set the current node as Trigger node.*
2. *Ignoring the void node, find the Best Forwarder based on Algorithm-1 for the Trigger node.*
3. *If the next hop node is the Destination node, send acknowledgement to the Trigger node. Exit.*
4. *If no forwarders are found, send Disrupt message to the Trigger node. Exit.*

5. If a forwarder is found, switch to Algorithm-1.
6. Exit.

## IV. RESULTS AND DISCUSSIONS

### A. Performance Evaluation

The performance of L-POR is evaluated through a simulation study using NS-2.34 [13]. Table 3 summarizes the simulation parameters. For simulation the network is modelled with several mobile nodes placed randomly. Both the protocols, Position-based Opportunistic Routing (POR) and L-POR are simulated independently and the performance metrics are evaluated.

TABLE 3
SIMULATION PARAMETERS

| Parameter | Value |
|---|---|
| Number of nodes | 160 |
| Transmission range | 225 m |
| Speed | 10, 30, 50, 100 m/s |
| Network topology | 800 x 800 m$^2$ |
| Antenna model | Omni antenna |
| Transmitter antenna gain | 1 dBi |
| Receiver antenna gain | 1 dBi |
| System loss factor | 1.0 |
| Transmitter signal power | 0.28 watts |
| Propagation model | Two-ray ground |
| Simulation time | 200 sec |

### B. Performance Metrics

- *Packet Delivery Ratio*: The ratio of the number of data packets received at the destination to the number of data packets sent by source.
- *End-to-end-delay*: The time taken for a packet to be transmitted from the source to the destination.
- *Path Length:* The average end-to-end number of hops for successful packet delivery.
- *Packet forwarding times per hop (FTH)*: The average number of times a packet is being forwarded to deliver a data packet over each hop given by,

$$FTH = \frac{N_s + N_f}{\sum_{i=1}^{N_r} N_{hi}} \quad (4)$$

where $N_s$, $N_f$, and $N_r$ are the number of packets sent at the source, forwarded at intermediate nodes, and received at the destination respectively. $N_{hi}$ is the number of hops for the i$^{th}$ packet that is successfully delivered.

- *Packet forwarding times per packet (FTP):* The average number of times a packet is being forwarded to deliver a data packet from the source to the destination given by,

$$FTP = \frac{N_s + N_f}{N_r} \quad (5)$$

### C. Comparative Analysis

The performance of L-POR is compared with POR. The performance of L-POR is evaluated.

The comparison graph for packet delivery ratio for the existing protocol, POR and the proposed protocol L-POR is shown in Fig 5. L-POR is found to achieve high packet delivery ratio, hence guarantees reliable packet delivery. The high ratio is due to the selection of most stable link as the next forwarder.

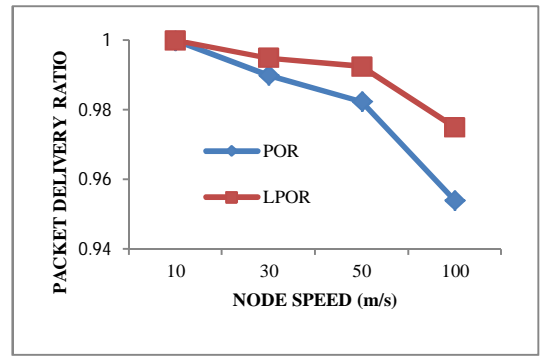

Fig. 5 Comparison graph for Packet Delivery Ratio

From Fig. 6 and 7, the packet forwarding times per hop and the number of forwarding times per packet are significantly reduced. The best forwarder is selected to be the node having maximum power of reception. Hence packet reforwarding is reduced considerably. Even if the best forwarder fails, the candidate nodes take over the forwarding function. Since these nodes are very close to the best forwarder, packet loss is eliminated. The candidate nodes can be accounted for such an improved performance.

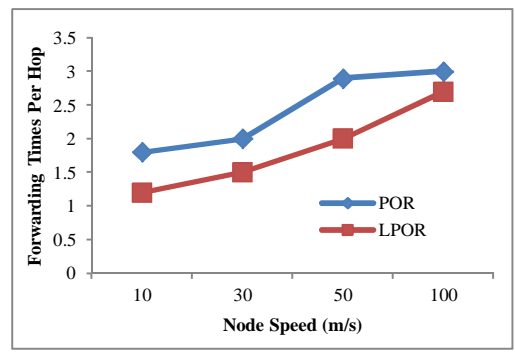

Fig. 6 Comparison graph for FTH

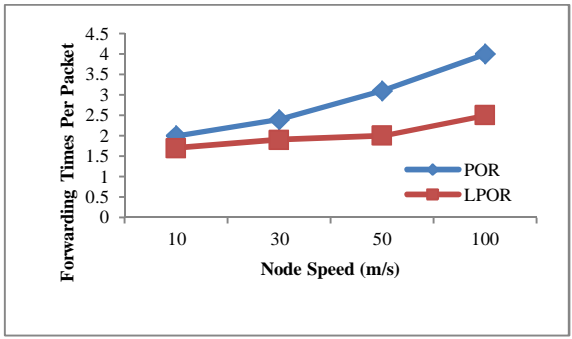

Fig. 7 Comparison graph for FTP

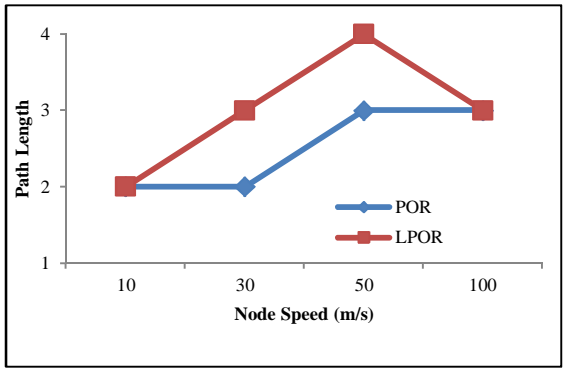

Fig. 8 Comparison graph for Path Length

Fig. 8 shows the comparison graph for path length. Unpredictable path length variation is found since the forwarder selection is not based on distance metric. Hence the hop count may not always be a minimal. This causes unpredictable end-to-end delay, as shown in Fig. 9.

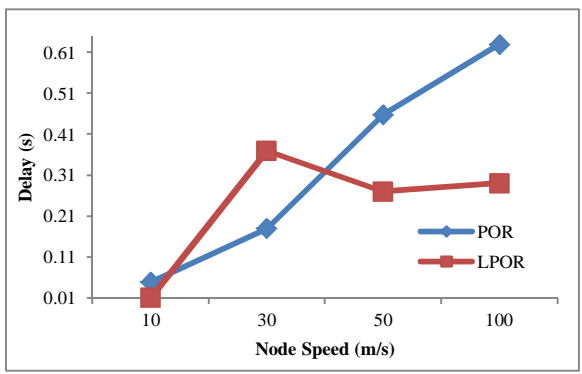

Fig. 9 Comparison graph for End-to-end delay

### V. CONCLUSION AND FUTURE WORK

A link based routing protocol L-POR is proposed for routing in MANET. The reception powers of all the one-hop neighbours that make positive progress towards the destination are calculated. The node with the highest power of reception is chosen as the best forwarder. To ensure better reliability, candidate nodes are also selected for each forwarder. In case the best forwarder fails, these candidate nodes take over the forwarding function according to their priorities. These nodes are selected to be nodes that lie closer to the best forwarder for better eavesdropping. Routing holes are also efficiently handled through an additional mechanism. This mechanism also avoids looping. Through simulation it is found that the packet delivery ratio of L-POR is better than that of POR.

L-POR guarantees reliability through best forwarder selection based on the node's link quality. Since the distance of the node towards the destination has not been considered for forwarder selection, the path length may not be always minimal causing a varying end-to-end delay. Hence future work can be done to reduce the hop count thus ensuring a lower end-to-end delay.

REFERENCES

[1] A. Trivino-Cabrera, S. Canadas-Hurtado, "Survey on Opportunistic Routing in Multihop Wireless Networks", *International Journal of Communication Networks and Information Security (IJCNIS),* Vol. 3, No. 2, August 2011

[2] Dazhi Chen, Pramod K. Varshney, "A Survey of void handling techniques for geographic routing in wireless networks", *IEEE communications survey*, 1st Quarter, Vol.9, No. 1, 2007

[3] Ana Maria Popescu, Ion Gabriel Tudorache, Bo Peng, A.H. Kemp, " Surveying Position Based Routing Protocols for Wireless Sensor and Ad-hoc Networks", *International Journal of Communication Networks and Information Security (IJCNIS)*, Vol. 4, No. 1, April 2012

[4] Rozner, E., Seshadri, J. , Mehta, Y. Lili Qiu, "SOAR: Simple Opportunistic Adaptive Routing Protocol for Wireless Mesh Networks", *IEEE Transactions on Mobile Computing,* Volume: 8 , Issue: 12, Page(s): 1622 - 1635 , Dec. 2009

[5] Jie Wu, Mingming Lu, and Feng Li, "Utility-Based Opportunistic Routing in Multi-hop Wireless Networks", *The 28th International Conference on Distributed Computing Systems,* Page(s): 470- 477 , June 2008

[6] Kai Zeng, Wenjing Lou, Jie Yang, Donald R. Brown III, "On Throughput Efficiency of Geographic Opportunistic Routing in Multihop Wireless Networks", *Springer, Mobile Networks Application,* Issue:12, Pages:347–357, 2007

[7] Luiz Carlos P. Albini, Antonio Caruso, Stefano Chessa and Piero Maestrini, "Reliable Routing in Wireless Ad Hoc Networks: The Virtual Routing Protocol", *Journal of Network and Systems Management,* Vol 14, No. 3, 2006

[8] Brad Karp, H. T. Kung," GPSR: Greedy Perimeter Stateless Routing for Wireless Networks", *Proceedings of the 6th Annual ACM/IEEE International Conference on Mobile Computing and Networking (MobiCom 2000),* 2000.

[9] Sanjit Biswas, Robert Morris, "ExOR: Opportunistic multi-hop routing for wireless networks", *Proceedings of the 2005 conference on Applications, technologies, architectures, and protocols for computer communications*, Pages 133-144 , 2005

[10] Jochen Schiller, *Mobile Communications* - Second edition, Pearson Education, Pg -36,41, 2009

[11] Stepanov, D. Herrscher, K. Rothermel, "On the impact of radio propagation models on MANET simulation results", *Proceedings of 7th International Conference on Mobile and Wireless Communications Networks (MWCN 2005), Marrakech, Morocco*, September 2005.
[12] Shengbo Yang, Chai Kiat Yeo, and Bu Sung Lee, "Toward Reliable Data Delivery for Highly Dynamic Mobile Ad Hoc Networks", *IEEE Transactions on Mobile Computing,* Vol. 11, No. 1, Jan 2012
[13] The Network Simulator NS-2, http://www.isi.edu/ns-nam/ns, 2011